\documentstyle[aps,prl,epsf,twocolumn]{revtex}
\begin{document}
\def\bea{\begin{eqnarray}}
\def\eea{\end{eqnarray}}
\def\a{\alpha}
\def\d{\delta}
\def\p{\partial} 
\def\nn{\nonumber}
\def\r{\rho}
\def\rv{\bar{r}}
\def\la{\langle}
\def\ra{\rangle}
\def\e{\epsilon}
\def\o{\omega}
\def\n{\eta}
\def\g{\gamma}
\def\break#1{\pagebreak \vspace*{#1}}
\draft
\title{Heat conduction in the disordered harmonic chain revisited}
\author{ Abhishek Dhar }
\address{ Raman Research Institute,
Bangalore 560080, India and Poornaprajna Institute, Bangalore.\\ }
\date{\today}
\maketitle
\widetext
\begin{abstract}
A general formulation is developed to study heat conduction in
disordered harmonic chains with arbitrary heat baths that satisfy the
fluctuation-dissipation theorem. A simple formal expression for the
heat current $J$ is obtained, from which its asymptotic system-size $(N)$
dependence is extracted. It is shown that the
``thermal conductivity'' depends not just on the 
system itself but also on the spectral properties of the
fluctuation and noise used to model the heat baths. 
As special cases of our heat baths we recover earlier results which
reported that for fixed boundaries $J \sim
1/N^{3/2}$, while for free boundaries  $J \sim 1/N^{1/2}$. 
For other choices we find that one can get
other power laws including the ``Fourier behaviour'' $J \sim 1/N$.
\end{abstract}

\pacs{PACS numbers: 44.10.+i, 05.70.Ln, 05.60.-k, 05.40.-a}
\narrowtext
The problem of heat conduction in one-dimensional classical systems of
interacting particles has attracted a lot of attention in recent
years \cite{bone}. A central issue here is determination of the 
dependence of the heat current $J$ on system size $N$.  According to
Fourier's law one expects $J \sim 1/N$ but 
a large number of studies
\cite{bone,lebo,rubin,cash,ishi,ding,FPU,toda,bambi,giard,prosen} suggest
that in one dimensions 
this may not always be true. Instead one finds that $J \sim 1/N^{\a}$
where $\alpha$ is usually different from one. 
Some obviously interesting and important 
questions are: what are the necessary and sufficient
conditions under which $\alpha=1$?,   what does $\a$ depend on and are
there universality classes? One of the main problems in the field has
been that most studies have been limited to numerical simulations of
nonlinear systems and it has been difficult to arrive at  definite
conclusions from  results of such studies. Thus, so far no clear
understanding has emerged.  We note that numerical simulations are
problematic because: (i) accurate numerical solutions of
nonlinear equations are very time consuming, (ii) equilibration
times are typically very long and this limits one to small 
system sizes and (iii) dependence on boundary conditions not clearly
understood. 

One of the earliest models to be investigated was the disordered
harmonic chain \cite{lebo,rubin,cash,ishi}. This problem is
analytically tractable to a large 
extent and the exponent $\a$ has been obtained analytically, though in
a semi-rigorous way. Surprisingly $\a$ seems to depend on boundary
conditions: for fixed boundary conditions (the Lebowitz model
\cite{cash}) $\a=1/2$, while for free boundaries (the Rubin-Greer model
\cite{rubin}) $\a=3/2$. This dependence on boundary conditions 
has not been understood in a precise way.

In this paper we revisit this problem. We present a general
formulation of the problem which enables one to view the two different
boundary conditions as two special cases of a range of possible
thermal reservoirs satisfying the fluctuation dissipation theorem.  
An approximate scheme, based on results from the theory of product of
random matrices, along with 
inputs from our numerical studies, enables us to obtain the asymptotic
\break{1.5in}
$N$-dependence of the current. 
We find 
the surprising result that {\it the exponent $\a$ depends not only
on the properties of the disordered chain itself, but also on the spectral
properties of the heat baths}. For special choices  of baths one
gets the ``Fourier behaviour''  $\a=1$.

We consider heat conduction through a one-dimensional disordered
harmonic chain. Particles $i=1,2...N$ with random masses 
are connected by harmonic springs with equal spring constants (set to
the value $1$). 
The Hamiltonian of the system is thus
\bea
H=\sum_{l=1}^N \frac{p_l^2}{2 m_l} 
+\sum_{l=0}^{N} \frac{(x_l-x_{l+1})^2}{2} 
\eea 
where $\{x_l\}$ are the displacements of the particles about their
equilibrium positions, $\{p_l\}$ their momenta and  $\{m_l\}$ are the
random masses. We put the boundary conditions $x_0=x_{N+1}=0$.
The particles in the bulk evolve through the classical equations of
motion while the boundary particles, namely particles $1$ and $N$ are
coupled to heat baths. The coupling to heat baths is effected by
including dissipative and noise terms in the equations of motion of
the end particles. The choice of the dissipative and fluctuating
forces is not unique. Different forms can be chosen provided 
that they satisfy the fluctuation-dissipation theorem.

We consider the following equations of motion for the particles:
\bea
m_1 \ddot{x_1} &=&-2 x_1+x_2 + \int_{-\infty}^t dt' A_L(t-t')
x_1(t')+\n_L(t) \nn \\
m_l \ddot{x_l} &=& - 2x_l+x_{l-1}+ x_{l+1}
~~~~~l=2,3...(N-1)  \nn \\
m_N \ddot{x_N} &=& -2 x_N+x_{N-1}  +
\int_{-\infty}^t d t'A_R(t-t') x_N(t')+\n_R(t), 
\label{eqmot}
\eea
where the terms $A_{L,R}(t)$ and $\n_{L,R}(t)$ describe dissipation
and noise, and 
will be specified later. We assume, unlike \cite{lebo}, that the heat
baths have been switched on at $t=-\infty$. 
To obtain the particular solution to these set of equations we define
the Fourier transforms 
$x_l(\o) = \int_{-\infty}^{\infty} x_l(t) e^{-i \o t}$;  
$\n_{L,R}(\o) = \int_{-\infty}^{\infty} \n_{L,R}(t) e^{-i \o t}$; 
$A_{L,R}(\o) = \int_{0}^{\infty} A_{L,R}(t) e^{-i \o t}$.
Plugging these into Eq.~(\ref{eqmot}) leads to the
following particular solution:
\bea
x_l(t) &=& \frac{1}{2 \pi} \int_{-\infty}^{\infty} d{\o}
\hat{Y}_{lm}^{-1}(\o) \hat{\n}_m(\o) e^{i \o t},~~~~~{\rm{where}}    
\label{solu} \\
\hat{Y} &=& \hat{\Phi}-\o^2 \hat{M}-\hat{A}(w) ~~~~~~ {\rm{with}}\nn \\
\hat{\Phi}_{lm} &=& - \d_{l,m+1}+2 \d_{l,m}-\d_{l,m-1} \nn \\
\hat{M}_{lm} &=& m_l \d_{l,m};~~~~\hat{A}_{lm}=\d_{l,m}(A_L(\o)
\d_{l,1}+A_R(\o) \d_{l,N}) \nn \\
\hat{\n}_l &=& \n_L(\o) \d_{l,1}+ \n_R(\o) \d_{l,N} \nn.
\eea
The full solution at time $t$ would be the sum of this particular
solution and a general solution of the homogeneous equation, which
would depend on the initial conditions. Since we are interested in the
steady state properties only, we will not require the general solution.

We now specify the properties of the dissipation and noise. 
Let us consider a system driven by a stationary noise $\n(t)$ with the
following correlator
\bea
\la \n(\o) \n(\o') \ra =2 \pi T I(\o) \d(\o+\o').
\label{noise}
\eea
If the dissipation is given by $A(\o)=a(\o)-i b(\o)$, where
$a(\o)$ and $b(\o)$ are real, then it follows
from the fluctuation dissipation theorem \cite{kubo} that the choice 
\bea
I(\o)= {2 b(\o)}/{\o}
\label{fdt}
\eea
ensures thermal equilibration of the system to the temperature $T$.
We choose  the same $I(\o)$ and $A(\o)$, satisfying Eq.~(\ref{fdt}), at
both boundaries. The noise correlators  given by Eq.~(\ref{noise}) are made 
different by setting $T=T_L$ at the left end and $T=T_R$ at the right end. 
For thermal equilibration it is necessary that the range of
frequencies, over which 
$I(\o)$ is non-zero, includes the normal modes of the
disordered chain, and we will only consider cases where this is true.  

For any given disorder realization, the energy current in the steady
state is given by   
\bea
J = \la [\int_{-\infty}^t dt'A_L(t-t') x_1(t')+\n_L(t)] \dot{x}_1(t) \ra ,
\eea
where $ \la...\ra $ denotes a noise average.
Using Eqs.~(\ref{solu},\ref{noise},\ref{fdt}),  and after some
algebraic manipulations, this reduces to the following simple form: 
\bea
J= \frac{T_L-T_R}{4 \pi} \int_{-\infty}^{\infty} d \o  t^2_N(\o)
~~~~{\rm{where}} \label{curr} \\
t^2_N(\o)=4 b^2(\o) Y^{-1}_{1N}(\o)   Y^{-1}_{1N}(-\o) \nn
\eea
We note that $t^2_N(\o)$, which is like a transmission coefficient,
depends both on the system and bath 
properties. We now proceed to write the current in a form where the
separate effects of the bath and 
system are more explicit. We first note that
\bea
&&Y^{-1}_{1N}(\o) Y^{-1}_{1N}(-\o) = |\Delta_N(\o)|^{-2}~~~
{\rm{with}}\nn \\ 
&& \Delta_N(\o) = Det[Y] \nn \\ 
&=& D_{1,N}-A(\o) (D_{2,N}+D_{1,N-1}) +A^2(\o)
D_{2,N-1} \nn \\ 
&=& \left( 1,~~-A(\o) \right) \left( \begin{array}{cc}
D_{1,N} & -D_{1,N-1} \\ D_{2,N} & -D_{2,N-1} \end{array} \right) \left( \begin{array}{c} 1 \\
A(\o)  \end{array}  \right),   
\label{expn}
\eea
where $D_{l,m}$ is defined to be  the determinant of
the submatrix of   $\hat{\Phi}-\o^2 \hat{M}$ beginning with the $l$th
row and column and ending with the $m$th row and column. Clearly $D_{l,m}$
depends on the system alone while $A(\o)$ depends on the bath. We
further note the following result which is easy to prove:
\bea
&& \left( \begin{array}{cc}
 D_{1,N} & -D_{1,N-1} \\ D_{2,N} & -D_{2,N-1} \end{array} \right)=T_1
T_2....T_N ~~~~~{\rm{where}} \label{dreln} \\
&& T_l=\left( \begin{array}{cc}
(2-m_l \o^2 & -1 \\ 1 & 0 \end{array} \right) \nn
\eea
The results of \cite{lebo,cash} follow from the following
choices of heat baths:
\bea
&& {\rm{(i)~ Lebowitz~model:}}~ A(\o)=-i \g \o;~I(\o)=2 \g,  \label{lebsp} \\
&& {\rm{(ii)~Rubin-Greer~model:}} \nn \\
&& A(\o)=1-\frac{\o^2}{2}-i \frac{\o}{2} {(4-\o^2)}^{1/2};~I(\o)= {(4-\o^2)^{1/2}}~~|\o|<2
\nn \\
&& A(\o)=1-\frac{\o^2}{2}+\frac{\o}{2} {(4-\o^2)}^{1/2};~I(\o)=0~~|\o|
>2 
\label{rubsp}
\eea 
Using these in Eq.~(\ref{curr}) we get the heat currents, $J_L$ and
$J_{RG}$, for the two models respectively as:
\bea
&& J_{L}=\pi^{-1}(T_L-T_R) \g^2 \int_{-\infty}^{\infty} d\o \o^2
j_N(\o)  \nn \\
&& j_N(\o)=  \{2 \g^2 \o^2 \nn \\ 
&& +D_{1,N}^2+\g^2 \o^2 (D_{1,N-1}^2+D_{2,N}^2)+ \g^4
\o^4 D_{2,N-1}^2 \}^{-1} \label{lebf}\\
&& J_{RG}=(4 \pi)^{-1} (T_L-T_R) \int_{-2}^{2} d\o \o^2
(4-\o^2) j_N(\o) \nn \\
&& j_N(\o)=  \{D_{1,N}^2+D_{2,N-1}^2  +(D_{1,N-1}+D_{2,N})^2 \nn \\
&&~~~~~~~~~~~ +2[2(1-\o^2/2)^2-1] D_{1,N} D_{2,N-1} \nn \\ &&  -2(1-\o^2/2) 
 (D_{1,N}+D_{2,N-1})(D_{1,N-1}+D_{2,N}) \}^{-1},
\label{rubf}
\eea
which are the same as in \cite{lebo,cash} (the differences are due to
a slightly different convention used by us).
Semi-rigorous arguments \cite{lebo,rubin,ishi} indicate that $\la
J_{L} \ra \sim 1/N^{3/2} $ 
while $ \la J_{RG} \ra \sim 1/N^{1/2} $, where the angular brackets now
denote a disorder average. For finite chains it is straightforward to
numerically compute the integrals appearing in
Eqs.~(\ref{lebf},\ref{rubf}) for given 
realizations of disorder and then perform disorder averages to obtain
$\la J_{L} \ra$ and  $ \la J_{RG} \ra$ . We show the results in
Fig.~(\ref{jvern}) for the case where the masses are chosen from a
uniform distribution between $1-\delta m$ to $1+ \delta m$. We do
get the expected power-law behaviours. 

We now present a scheme which allows us to determine
the $N$-dependence of the current for arbitrary choices of heat
baths. This is based on the following observations: 

(i) The first observation follows from the Furstenberg theorem on
the limiting form of product of random noncommuting variables. For the case
considered here, the theorem states that, for almost any choice
of the sequence of random masses $\{ m_l \}$,
\bea 
\lim_{N \to \infty} \frac{1}{N} \log | T_1 T_2....T_N u |= \g(\o) > 0   
\label{furst}
\eea  
for any non-zero vector $u$, with fluctuations of order
$1/\sqrt{N}$. Further it can be shown that 
\cite{lebo,ishi} in the limit $\o \to 0^{+}$,
\bea
 \g( \o ) \to {(<m^2>-<m>^2)}w^2/{(8 <m>)}. 
\label{glim}
\eea
This means [from Eq.~(\ref{dreln})] that the $D_{l,m}$, which occur in
the denominator of the integrand in Eq.~(\ref{curr}), diverge
exponentially with $N$, and hence the only significant contribution to the current
comes from low frequency components of order $\stackrel{<}{\sim} 1/N^{1/2}$. 
We note that the fact that low frequency modes are extended follows
from the translational invariance of the random-mass model.

(ii) The result Eq.~(\ref{glim}) has been obtained in the strict limit
of $N \to \infty$ when the ratio of successive particle displacements
reaches a stationary state. For finite $N$, we find from our numerical
studies that this result is true only for $\o
\stackrel{>}{\sim} 1/N^{1/2}$. In Fig.~(\ref{growth}) we have plotted
$\la |D_{1,N}| \ra $ as a function of frequency. We find that the exponential
growth predicted by Eq.~(\ref{glim}) does not occur at $\o
\stackrel{<}{\sim} 1/N^{1/2}$. In this range we find instead [see
Fig.~(\ref{smallw})] that
$D_{1,N}$ is very accurately given by its form for the
ordered case with masses all equal to $<m>=1$.
Thus over the range $\o \stackrel{<}{\sim} 1/N^{1/2}$, we 
shall approximate $j(\o)$ in Eq.~(\ref{curr}) by its form 
for the ordered chain. We expect this approximation to be good as
long as we are interested only in the asymptotic $N$-dependence. 

For the equal mass case
one has $D_{1,N}=\sin{[k(N+1)]}/\sin{(k)}$ where $\o=2 \sin{(k/2)}$. 
Hence within our approximate scheme we then get the following
expression for the disorder-averaged current: 
\bea
&&\la J \ra \sim  (T_L-T_R) \int_{0}^{1/N^{1/2}} f(k) dk \label{currapp} \\ 
&& f(k)={b^2(\o) \sin^2(k) \cos(k/2)} \times  \{ |\sin[k(N+1)] \nn \\
&& ~~~~~~~~~-2A(\o) \sin(kN) +A^2(\o) \sin[k(N-1)]|\} ^{-2}. \nn
\eea
It is clear that the form
of $A(\o)$ at low frequencies will determine the asymptotic
$N$-dependence of the current. For the Lebowitz model $A(\o) = -i \g \o
$ while for the Rubin-Greer model $A(\o) \sim 1-i \o$ and
Eq.~(\ref{currapp}) does give the expected $1/N^{3/2}$ and $1/N^{1/2}$
behaviour for the two cases respectively. 
In general we find that $J \sim \frac{1}{N^{\a}}$ where the exponent $\a $
depends on the low-frequency behaviour of $A(\o)$. Some special cases
are: 

\noindent (i) $A(\o) \sim -i sgn(\o)\o^s$: Eq.~(\ref{currapp}) then gives
$\a=s/2+1$ for $s >0$.

\noindent (ii) $A(\o) \sim 1-i sgn(\o) \o^s$: in this case we get
$\a=1-s/2$ for $ 0<s<1$ and $\a=s/2$ for $s \ge 1$. Note that the case $s=2$
gives $\a=1$ that is, a Fourier-like behaviour. We verify this by a
numerical evaluation of the integral in Eq.~(\ref{curr}) for
chains of finite length and given disorder, and then averaging of the
current over many 
disorder realizations. The result is shown in Fig.~(\ref{jvern}).

One simple way of generating thermal sources with different spectral
properties is to couple the disordered chain to an infinite set of
oscillators in thermal equilibrium. The distribution of oscillator
frequencies can be arbitrary except that it should include the range
of the disordered chain frequencies. In this case it can be shown that
the equations of motion are of the general form Eq.~(\ref{eqmot}) with
$A(t)= \int_0^{\infty} G(\o_q) \sin{(\o_q t)} d\o_q$ where $G(\o_q)$
depends on the choice of oscillator frequencies. The Rubin-Greer
model, where the 
bath is simply an infinite ordered chain, corresponds to the choice
$G(\o_q)=\frac{1}{\pi} \o_q (4-\o_q^2)^{1/2}$ for $\o_q \leq 2$ and zero
elsewhere. 

Finally, we have also studied the effect of introducing a quadratic
external potential, in addition to the mass disorder. In this case, the
low frequency current-carrying 
modes are suppressed and we find that the current decays exponentially
with system-size.   

In summary we have studied the nonequilibrium steady
state of a mass disordered harmonic chain coupled to heat
baths at different temperature. We have shown that the system size
dependence of the energy current, given by $J \sim 1/N^{\a}$, is
determined 
not just by the properties of the system itself but also by those of
the heat baths. One gets a continuous set of exponents $\a$ depending
on the low frequency spectral properties of the bath. This
seems contrary to the general belief in nonequilibrium statistical mechanics
that the steady state of a close-to-equilibrium system will not depend
on the details of the boundary conditions sustaining the steady state.    
We explain this by arguing as follows: the integrability of
the harmonic system prevents local thermal equilibrium from being
established and so the system is really far-from-equilibrium. 
Each of the noninteracting modes independently carries some energy
current and the total current 
depends on how exactly the heat baths distribute energy among the
various modes.  
For non-integrable systems, we expect that there should be transfer of
energy amongst various modes leading to a state of 
local thermal equilibrium. Hence energy transport should be
independent of boundary conditions. However  careful studies are needed to
verify that this is actually true, especially for systems which may be
close to integrable ones. Indeed the possibility of boundary
condition dependence is suggested from the results of some studies on nonlinear
models \cite{FPU,toda}. 
These have often been attributed to the fact that  some boundary conditions
lead to jumps in the temperatures at the boundaries which in turn 
makes it difficult to extract the correct system-size dependence of
current. However our study  shows the possibility of
boundary-condition dependence even in the absence of such jumps. 

Some other interesting questions are: (1) are
the peculiarities of the harmonic chain  generic to any integrable
system, (2) are these results also true for harmonic systems in higher
dimensions and (3) can the present formalism be extended to the
quantum-mechanical case. The answers to these questions may have
implications for understanding experiments on heat conduction in
systems such as insulating nanowires,  
where similar boundary related effects could lead to modification of
Landauer-type formulas for thermal conductivity\cite{meso}.  

I thank Madan Rao for very helpful discussions.

\vbox{
\vspace{1.0cm}
\epsfxsize=8.0cm
\epsfysize=6.0cm
\epsffile{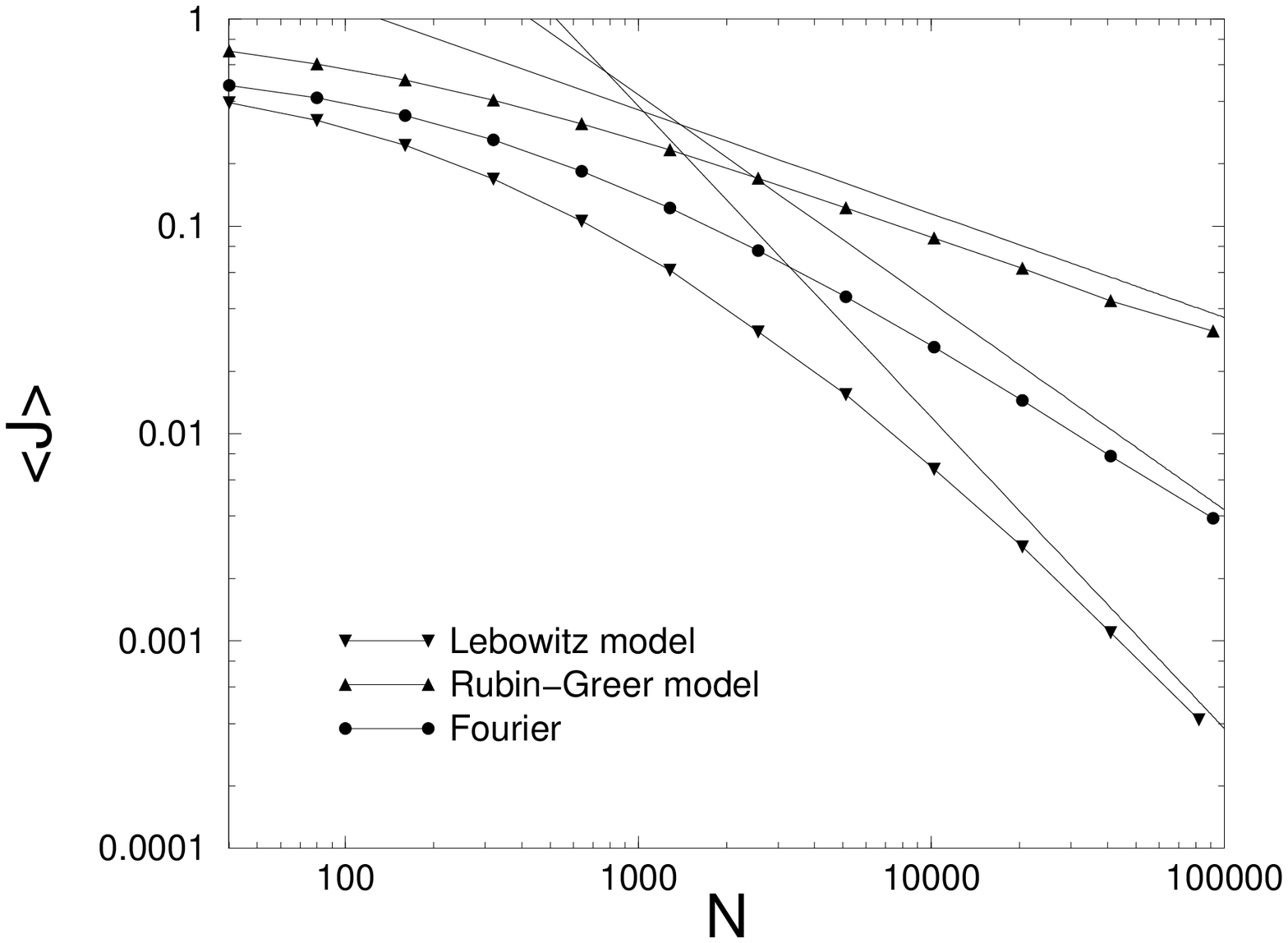}
\begin{figure}
\caption{\label{jvern} 
System size dependence of the disorder averaged steady state current
for three different models of heat baths. The straight lines shown
have slopes $1/2$, $1$ and $3/2$. In all cases the disorder strength
$\delta m=0.2$. The error in the measurements is of the order of the
size of the points. }
\end{figure}}
\vbox{
\epsfxsize=8.0cm
\epsfysize=6.0cm
\epsffile{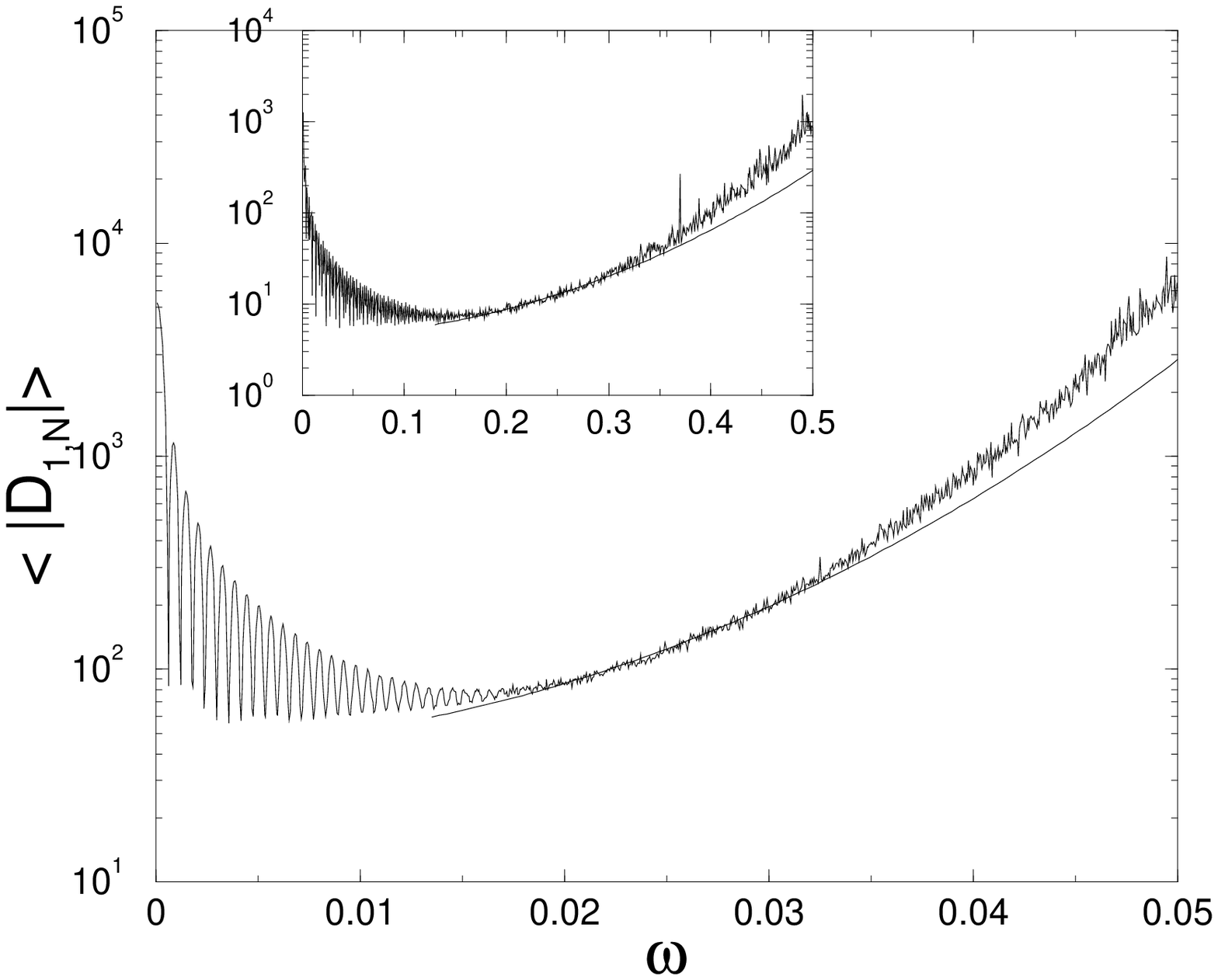}
\begin{figure}
\caption{\label{growth} Growth of solutions in a random harmonic chain for
$N=10^6$ and $N=10^4$ (inset). The disorder strength was taken to be
$\delta m=0.2$. Note that the exponential growth starts from $\o \approx  
c/\sqrt{N}$ (with $c \approx 13$). The smooth solid curves correspond to the exponential
growth predicted by Eq.~(\ref{glim}).} 
\end{figure}}
\vbox{
\epsfxsize=8.0cm
\epsfysize=6.0cm
\epsffile{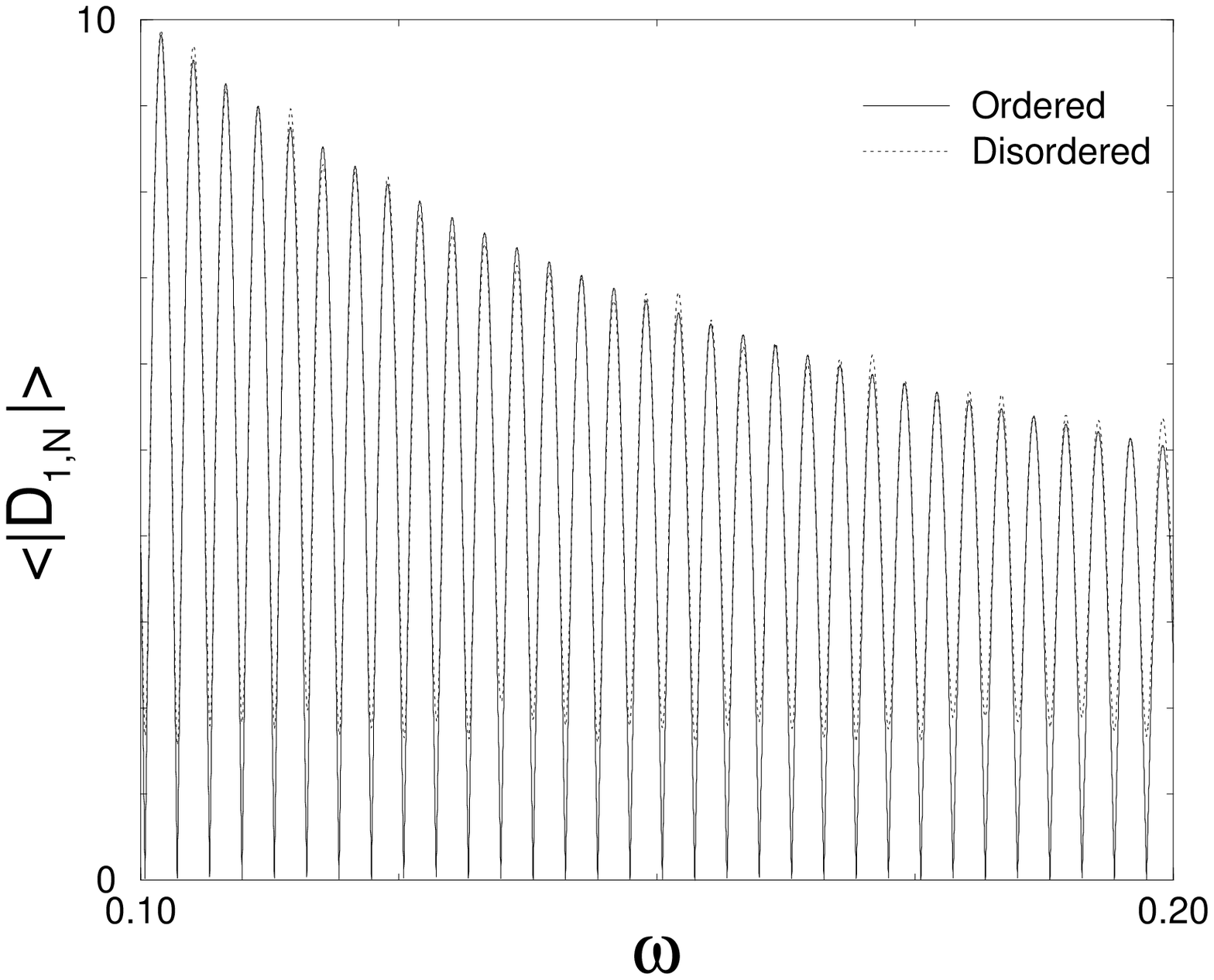}
\begin{figure}
\caption{\label{smallw} Frequency dependence of $\la |D_{1,N}| \ra$ at
small $w$ for $\delta m=0.2$ and $N=10^4$ is compared with $|D_{1,N}|$ for the ordered chain.  }
\end{figure}}
\end{document}